\documentclass[10pt]{article}
\usepackage{latexsym}
\usepackage{amsmath}
\usepackage{pstricks} 
\usepackage{indentfirst}

\long\def\ca#1\cb{} 

\newcommand{\ad}{^\dagger }

\newcommand{\AND}{\mbox{\small AND}}

\newcommand{\avg}[1]{\langle #1\rangle }
\newcommand{\becs}{\begin{cases}}
\newcommand{\bem}{\begin{matrix}}

\newcommand{\bra}[1]{\langle#1|}

\newcommand{\dg}{^\circ } 
\newcommand{\dya}[1]{|#1\rangle\langle#1|}

\newcommand{\encs}{\end{cases}}
\newcommand{\enm}{\end{matrix}}

\newcommand{\hf}{{\textstyle\frac{1}{2} }}

\newcommand{\hquad}{\mspace{8 mu}} 

\newcommand{\inpd}[2]{\langle#1|#2\rangle }
\newcommand{\ket}[1]{|#1\rangle }

\newcommand{\lra}{\leftrightarrow }

\newcommand{\mted}[3]{\langle#1|#2|#3\rangle }

\newcommand{\NOT}{\mbox{\small NOT}}
\newcommand{\od}{\odot }
\newcommand{\OR}{\mbox{\small OR}}
\newcommand{\ot}{\otimes }

\newcommand{\ra}{\rightarrow }
\newcommand{\Ra}{\Rightarrow }

\newcommand{\st}{\sqrt{2}}
\newcommand{\tm}{\times }

\newcommand{\vb}{\,|\,}

\newcommand{\DC}{{\mathcal D}}

\newcommand{\FC}{{\mathcal F}}

\newcommand{\HC}{{\mathcal H}}

\newcommand{\PC}{{\mathcal P}}
\newcommand{\QC}{{\mathcal Q}}

\newcommand{\al}{\alpha }
\newcommand{\bt}{\beta }

\newcommand{\ep}{\epsilon}

\newcommand{\zt}{\zeta }

\newcommand{\kp}{\kappa }

\def\outl#1{\par{\medskip\noindent\hspace*{.5cm}\bf
      \mathversion{bold}#1\mathversion{normal}\smallskip} }
 \def\xa{} \def\xb{}  

 \def\outl#1{}  \def\xa{} \def\xb{}  

\ca
 \def\outl#1{\par{\medskip\noindent\hspace*{.5cm}\bf
      \mathversion{bold}#1\mathversion{normal}\smallskip} }
 \long\def\xa#1\xb{}
 
\cb

\begin{document}

\title{Particle Path Through a Nested Mach-Zehnder Interferometer}
\author{Robert B. Griffiths\thanks{Electronic address: rgrif@cmu.edu}\\
Department of Physics\\
Carnegie Mellon University\\
Pittsburgh, PA 15213}
\date{Version of 1 July 2016}
\maketitle

\begin{abstract}
  Possible paths of a photon passing through a nested Mach-Zehnder
  interferometer on its way to a detector are analyzed using the consistent
  histories formulation of quantum mechanics, and confirmed using a set of weak
  measurements (but not weak values). The results disagree with an analysis by
  Vaidman [ Phys.\ Rev.\ A 87 (2013) 052104 ], and agree with a conclusion
  reached by Li et al.\ [ Phys.\ Rev.\ A 88 (2013) 046102 ]. However, the
  analysis casts serious doubt on the claim of Salih et al.\ (whose authorship
  includes Li et al.) [ Phys.\ Rev.\ Lett.\ 110 (2013) 170502 ] to have
  constructed a protocol for counterfactual communication: a channel which can
  transmit information even though it contains a negligible number of photons.
\end{abstract}

\tableofcontents

\section{Introduction}
\label{sct1}

\xb
\outl{Path of particle thru nested MZI. Connection with second
  measurement problem}
\xa

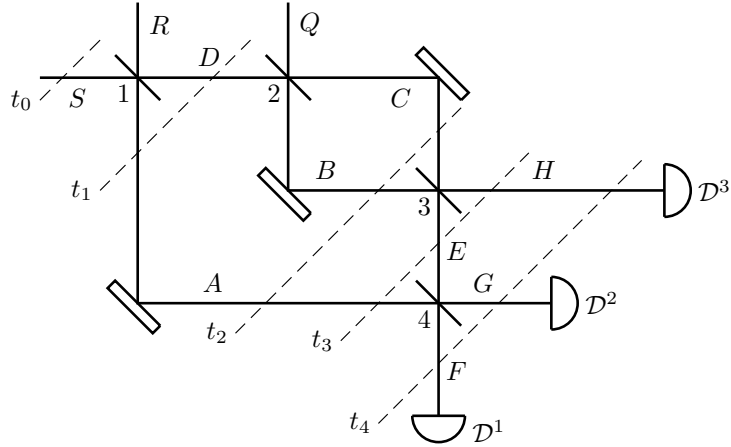
\begin{figure} [h]
$$
\begin{pspicture}(-1,-1.5)(6.5,4.5) 
\newpsobject{showgrid}{psgrid}{subgriddiv=1,griddots=10,gridlabels=6pt}
\def\lwd{0.035} 
\def\lwb{0.10}  
\def\lwn{0.015}  
\psset{
labelsep=2.0,
arrowsize=0.150 1,linewidth=\lwd}
\def\rdet{0.35}  
\def\detect{
\psarc[fillcolor=white,fillstyle=solid](0,0){\rdet}{-90}{90}
\psline(0,-\rdet)(0,\rdet)}
\def\drad{0.35}
\def\detectd{
\psarc[fillcolor=white,fillstyle=solid](0,0){\drad}{180}{0}
\psline(-\drad,0)(\drad,0) }
\def\bsw{0.3}
\def\bsp{\psline[linewidth=\lwd](-\bsw,\bsw)(\bsw,-\bsw)}
\def\mrw{0.35}
\def\mrw{0.3}\def\mrwm{0.2}\def\mrwp{0.4}
\def\mirrorv{
\psline[linewidth=\lwd]%
(\mrw,-\mrw)(\mrwm,-\mrwp)(-\mrwp,\mrwm)(-\mrw,\mrw)(\mrw,-\mrw)}
\def\mirrore{
\psline[linewidth=\lwd]%
(\mrw,-\mrw)(\mrwp,-\mrwm)(-\mrwm,\mrwp)(-\mrw,\mrw)(\mrw,-\mrw)}
\psline(-1.3,3)(4,3)(4,-1.5)
\psline(0,4.0)(0,0)(5.5,0)
\psline(2,4)(2,1.5)(7,1.5)
\rput(0,3){\bsp} \rput(2,3){\bsp} \rput(4,1.5){\bsp} \rput(4,0){\bsp}
\rput(0,0){\mirrorv} \rput(2,1.5){\mirrorv} \rput(4,3){\mirrore}
\rput(7,1.5){\detect} \rput(5.5,0){\detect} \rput(4,-1.5){\detectd}
\psline[linestyle=dashed,linewidth=\lwn](-1.3,2.7)(-0.5,3.5)
\rput[r](-1.4,2.7){$t_0$}
\psline[linestyle=dashed,linewidth=\lwn](-0.5,1.5)(1.5,3.5)
\rput[r](-.6,1.5){$t_1$}
\psline[linestyle=dashed,linewidth=\lwn](1.3,-0.4)(4.3,2.6)
\rput[r](1.2,-0.4){$t_2$}
\psline[linestyle=dashed,linewidth=\lwn](2.7,-0.5)(5.2,2.0)
\rput[r](2.6,-0.5){$t_3$}
\psline[linestyle=dashed,linewidth=\lwn](3.2,-1.6)(6.7,1.9)
\rput[r](3.1,-1.6){$t_4$}
\rput[l](0.15,3.7){$R$} \rput[l](2.15,3.7){$Q$}
\rput[t](-0.8,2.85){$S$} \rput[b](0.95,3.15){$D$} \rput[l](4.1,0.7){$E$}
\rput[b](1,0.15){$A$} \rput[b](2.5,1.65){$B$} \rput[t](3.5,2.85){$C$}
\rput[b](5.4,1.65){$H$} \rput[b](4.6,0.15){$G$} \rput[l](4.1,-0.9){$F$}
\rput[l](4.45,-1.7){$\DC^1$} 
\rput[l](5.95,0){$\DC^2$} 
\rput[l](7.45,1.5){$\DC^3$}
\rput[rt](-.1,2.9){1} \rput[rt](1.9,2.9){2} 
\rput[rt](3.9,1.4){3} \rput[rt](3.9,-0.1){4}
\end{pspicture}
$$
\caption{%
  Nested Mach-Zehnder interferometer (MZI). The tilted solid lines are beam
  splitters numbered 1, 2, 3, 4; the double tilted lines are mirrors; the
  semicircles are detectors. The horizontal and vertical lines indicate
  different channels which are possible particle (photon) paths. The
  reflectivities and phases of beam splitters 2 and 3 associated with the inner
  MZI are chosen so that a particle entering through $D$ will exit through $H$
  and be detected by $\DC^3$, rather than passing into $E$. The intersections
  of the dashed lines with the particle paths indicate possible locations of
  the particle at the successive times $t_0<t_1<t_2<t_3<t_4$. }
\label{fgr1}
\end{figure}

This article addresses the question of what one can say about the path of a
photon, hereafter called a ``particle'', as it passes through the nested
Mach-Zehnder interferometer (MZI) shown in Fig.~\ref{fgr1} on its way to one of
the three detectors. This setup is of interest in and of itself because it
raises a question that cannot be answered by standard quantum mechanics as
found in standard textbooks: what is a microscopic quantum system actually
doing prior to measurement by a macroscopic apparatus? Research in quantum
foundations has yet to supply any widely accepted answer to the infamous
\emph{measurement problem}: provide a consistent, fully quantum mechanical
description of the entire process that goes on in an actual physical
measurement of a microscopic system. Indeed, even the \emph{first} measurement
problem, understanding how such a measurement can have a well-defined outcome
or pointer position, to use the archaic but picturesque language of quantum
foundations, rather than a quantum superposition of different and
macroscopically distinct positions, has given rise to a long and inconclusive
discussion. The failure to settle this first problem has diverted attention
from the equally important \emph{second} measurement problem \cite{Grff15}: how
to infer (or retrodict) the earlier microscopic property, the one the apparatus
was designed to measure, from the final pointer position. Physicists who do
experiments frequently interpret the outcomes using realistic language such as:
``the detector was triggered by a fast muon traveling from the region where the
protons collided.'' If quantum theory cannot, at least in principle, make sense
of language of this sort, how can one claim that experiment has confirmed what
is often said to be a very successful physical theory?

\xb
\outl{Vaidman vs. Li et al.\ \& Salih et al.; counterfactual communication}
\xa

In addition to its intrinsic interest, the gedanken experiment of
Fig.~\ref{fgr1} is central to an ongoing disagreement between Vaidman, who
analyzed it in \cite{Vdmn13}, and Li et al., who reached a different conclusion
in \cite{LAAZ13}, to which Vaidman replied in \cite{Vdmn13b}. Around the same
time Salih et al. \cite{Slao13}---the authorship includes that of
\cite{LAAZ13}---claimed to have invented a quantum protocol capable of
\emph{counterfactual communication}: messages can be sent from Bob to Alice
through a communication channel that contains a negligible number of photons
(particles), indeed zero in an ideal asymptotic limit. This protocol is an
extended and more complicated version of Fig.~\ref{fgr1}, with many successive
beam splitters and mirrors (or repeated passes through and reflections from a
small number of beam splitters and mirrors). The claim that it achieves
counterfactual communication was challenged by Vaidman in \cite{Vdmn14}, with a
response by Salih et al.\ in \cite{Slao14}. For an extensive bibliography,
including references to some experiments, see \cite{Vdmn15,Vdmn15b}.

\xb
\outl{This paper: CH used to analyze nested MZI. Previous studies have problems}
\xa

The present paper discusses the gedanken experiment in Fig.~\ref{fgr1} using
the \emph{consistent histories} (CH), also known as the decoherent histories,
formulation of quantum theory, which unlike standard quantum mechanics does not
treat measurement as an unanalyzable primitive concept, but instead as an
example of a quantum physical process governed by exactly the same fundamental
principles that apply to all such processes. The CH approach is internally
consistent (does not lead to unresolvable paradoxes) and makes the same
predictions for macroscopic measurement outcomes, using much the same
mathematics, as do the textbooks. See Sec.~\ref{sct3} below for further remarks
and some references. In the case of the nested MZI in Fig.~\ref{fgr1} with
particular reference to a particle detected by $\DC^1$, the CH study leads to
the result that previous analyses, while correct in certain respects, have made
assumptions which are not fully consistent with quantum principles. Hence both
the claim of counterfactual communication and Vaidman's criticism thereof have
serious deficiencies.

\xb
\outl{Structure of this article}
\xa

The structure of the remainder of this article is as follows:
Section~\ref{sct2} contains details of the nested MZI and the competing claims
about the path followed by a particle (photon) passing through it on its way to
the $\DC^1$ detector. Possible paths are analyzed in Sec.~\ref{sct3} using
consistent histories, while Sec.~\ref{sct4} has additional remarks which may
assist readers unfamiliar with the histories approach. The results in
Sec.~\ref{sct3} are consistent with a study in Sec.~\ref{sct5} using weak
measurements, and are compared with Vaidman's use of the two state vector
formalism in Sec.~\ref{sct6}. Section~\ref{sct7} discusses what appears to be a
serious difficulty with the claim of counterfactual communication. The
conclusions are summarized in Sec.~\ref{sct8}.

\section{Nested Mach-Zehnder Interferometer}
\label{sct2}

\subsection{Interferometer and beam splitters}
\label{sbct2.1}

\xb
\outl{Figure showing nested MZI. Beamsplitters, mirrors, detectors}
\xa

In the nested Mach-Zehnder interferometer (MZI) shown in Fig.~\ref{fgr1}, a
particle from a source channel $S$ enters the outer MZI at beam splitter 1 and
can pass through the lower arm $A$, via a mirror, to beam splitter 4. However,
the upper arm of the outer MZI is interrupted by an inner MZI, whose arms $B$
and $C$, between beam splitters 2 and 3, connect the input channels $D$ and $Q$
to the output channels $E$ and $H$, with $E$ going on to beam splitter 4. The
reflectivities and phases of beam splitters 2 and 3 associated with the inner
MZI are chosen so that a particle entering through $D$ will exit through $H$
and be detected by $\DC^3$, rather than passing into $E$. The output ports $F$
and $G$ of beam splitter 4 lead to detectors $\DC^1$ and $\DC^2$, respectively.
(The detectors are denoted by a different font and labeled with superscripts to
avoid any confusion with the $D$ channel and with the ket $\ket{D_1}$ in
\eqref{eqn1}.) The dashed lines in the figure indicate possible locations of
the particle at successive times $t_0 < t_1 < t_2 < t_3 < t_4$ on its journey
from $S$ at $t_0$ to one of the detectors.

\xb
\outl{Unitary transformations associated with beam splitters}
\xa

Let $T_{kj}$ be the unitary time transformation that
maps a ket at time $t_j$ to its counterpart at $t_k$. Here is a suitable
collection of unitaries (the choice is not unique); $T_{j,j-1}$ represents the
action of beam splitter $j$ in Fig.~\ref{fgr1}:
\begin{align}
&T_{10}:\; 
\ket{S_0} \ra \al\ket{A_1} + \bt\ket{D_1},\quad
\ket{R_0} \ra -\bt \ket{A_1} + \al\ket{D_1},\quad
\ket{Q_0}\ra \ket{Q_1};\quad
\notag\\
&T_{21}:\; \ket{A_1} \ra \ket{A_2},\quad 
\ket{D_1} \ra r\ket{B_2} + r\ket{C_2},\quad 
\ket{Q_1} \ra r\ket{B_2} - r\ket{C_2};
\notag\\
&T_{32}:\; \ket{A_2} \ra \ket{A_3},\quad
 \ket{B_2} \ra -r\ket{E_3} + r\ket{H_3},\quad 
\ket{C_2} \ra r\ket{E_3} + r\ket{H_3};
\notag\\
&T_{43}:\;\ket{A_3} \ra \al\ket{F_4} + \bt\ket{G_4},\quad
\ket{E_3} \ra \bt\ket{F_4} -\al\ket{G_4},\quad
\ket{H_3} \ra \ket{H_4}.
\label{eqn1}
\end{align}
The coefficients are real numbers:
\begin{equation}
 0 < \al,\bt < 1, \text{ with } \al^2 + \bt^2 = 1;\quad r:=1/\st.
\label{eqn2}
\end{equation}
The unitaries for larger time steps can be obtained using $T_{lj} =
T_{lk}T_{kj}$, and for inverse time steps using $T^{}_{jk} = T_{kj}\ad$.
One should think of kets such as $\ket{A_1}$ or $\ket{B_2}$ as normalized wave
packets located in the corresponding channels near the points where the dashed
time lines cross the straight lines indicating the channels, and mutually
orthogonal since they are far apart. In all there are fifteen of these kets in
\eqref{eqn1}. While in principle each $T_{j,j-1}$ should tell what
happens to every one of them, \eqref{eqn1} supplies all the information about
unitary time development that is needed for the present discussion. We are
only interested in the situation in which the particle starts off in
$\ket{S_0}$ at $t_0$, and is in either $\ket{A_1}$ or $\ket{D_1}$ at time
$t_1$, etc. 

\subsection{Where was the particle?}
\label{sbct2.2}

\xb \outl{Thru what path did particle arrive at $\DC^1$?. Unitary evolution of
  $\ket{\psi_0}$} \xa

Suppose a particle that enters the outer MZI from $S$ is later detected by
$\DC^1$. Where was it at earlier times $t_1$, $t_2$, and $t_3$, when it was
still inside the interferometer? Readers are invited to work out their own
answers before considering those discussed below. Here is the succession of
states $\ket{\psi_j} = T_{j0}\ket{\psi_0}$ obtained by unitary time evolution
using the transition amplitudes in \eqref{eqn1}, starting with $\ket{\psi_0} =
\ket{S_0}$:
\begin{align}
\ket{\psi_1} = \al\ket{A_1} + \bt\ket{D_1}, &\quad
\ket{\psi_2} = \al\ket{A_2} + r\bt \bigl(\,\ket{B_2} +\ket{C_2}\bigr),
\notag\\
\ket{\psi_3} = \al\ket{A_3} + \bt\ket{H_3},&\quad
\ket{\psi_4} = \al^2 \ket{F_4} + \al\bt\ket{G_4} + \bt\ket{H_4},
\label{eqn3}
\end{align}
where $\ket{\psi_j}$ corresponds to the time $t_j$ in Fig.~\ref{fgr1}.

\xb
\outl{Li et al.: particle was in $A$. This not a consequence of
std QM with wavefn collapse}
\xa

Li et al.\ in \cite{LAAZ13} (and Salih et al. in \cite{Slao13,Slao14}) assert
that if $\DC^1$ is triggered, then the particle was in channel $F$ at $t_4$ and
in channel $A$ at the earlier times $t_1$, $t_2$, and $t_3$. In Sec.~\ref{sct3}
we will show that this is, in a sense which will be made clear, a correct
answer. However, contrary to the claim in \cite{Slao14}, it is \emph{not} a
consequence of ``standard quantum mechanics,'' if by that one means what is
found in standard textbooks. A student taking the final examination in Quantum
101 knows only that the quantum wave function develops unitarily in time until
a measurement is made, leading to a mysterious collapse. And since at none of
the times after $t_0$ and before the final measurement (at time $t_5$) is
$\ket{\psi_j}$ confined to a single channel, the question of the particle's
location cannot be answered using the textbook approach.

\xb
\outl{SQM: black box links preparation, measurement. Must open the box}
\xa

In practice, using quantum mechanics often requires going beyond, and sometimes
forgetting, what is taught in textbooks, especially if one wants to take the
talk of experimentalists seriously. Standard quantum mechanics as applied to
the situation under discussion is best thought of as a series of rules for
calculating the probabilities of certain macroscopic measurement outcomes given
an earlier preparation, with $\ket{\psi(t)}$ at intermediate times
a useful calculational tool, but otherwise not interpreted. This is a ``black
box'' approach in that we (think we) understand preparations and measurements,
while what goes on in between, inside the black box, is best not discussed.
Those who dare open the box to try and figure out what is happening inside it
risk going insane.%
\footnote{A paraphrase of Feynman, p.~129 of \cite{Fynm65}. } %
But if we are to assess the claims and counterclaims mentioned in
Sec.~\ref{sct1} we must open the box.

\xb
\outl{Plausible line of reasoning leading to `particle was in $A$'}
\xa

One line of reasoning which sounds plausible and will lead to the conclusion
given in \cite{LAAZ13} can be worded as follows. The particle after it passed
through beam splitter 1 was either in arm $A$ or in arm $D$. If it was in arm
$D$, then because of the action of beam splitters 2 and 3 we know that it would
have emerged in $H$, not in $E$. Had it emerged in $H$, then $\DC^3$ would have
detected it, but since it was detected by $\DC^1$ it was not detected by
$\DC^3$. Thus at $t_1$ the particle was surely \emph{not} in channel $D$, and
hence it must have been in $A$, and as there is no way the particle could
escape from $A$ before reaching beam splitter 4, it must have been in $A$ at the
times $t_1$, $t_2$, and $t_3$.

\xb
\outl{Preceding reasoning a precarious route to a correct conclusion}
\xa

Such reasoning has, as we shall see, arrived at a correct conclusion, but by a
precarious route that is not altogether convincing. To begin with, what
justifies treating $\al\ket{A} + \bt\ket{D}$ as ``the particle was either in
$A$ or it was in $D$''? Interpreting superpositions in this way is hazardous,
and fails badly when applied to the double slit paradox. Next come some ``if
... then'' counterfactuals,%
\footnote{Not to be confused with `counterfactual' as used in `counterfactual
  communication.'} %
and counterfactual reasoning in a quantum context has many pitfalls; see, for
example, the exchange in \cite{Stpp12,Grff12b}.

\xb
\outl{Vaidman: Li et al.\ ``a naive classical approach''; Vd prefers TSVF, weak
  measurements}
\xa

\xb \outl{TSVF phenomenology: particle was where both $\{\bra{\phi_j}\}$,
  $\ket{\psi_j}$ nonzero; $\{\bra{\phi_j}\}$ for $\bra{\phi_4} = \bra{F}$} \xa

So one can understand why Vaidman \cite{Vdmn14} dismisses arguments of this
sort as ``a naive classical approach.'' His own analysis in \cite{Vdmn13}
reaches a different conclusion through employing the two state vector formalism
(TSVF) \cite{AhVd90,Vdmn10}, plus some consideration given to weak and to
strong nondestructive measurements. Let us begin with the TSVF; weak
measurements will be discussed in Sec.~\ref{sbct5.2}. Along with the usual
`forwards' wave advancing unitarily in time from an earlier preparation, the
$\ket{\psi_j}$ in \eqref{eqn3}, the TSVF employs a wave moving `backwards' from
a later measurement or, to be more precise, a property revealed by a later
measurement. In the case at hand the later measurement result is detection by
$\DC^1$, the corresponding property is $\ket{F}$, and one defines
\begin{equation}
 \ket{\phi_j} = T_{j4} \ket{F_4}, \text{ or } \bra{\phi_j} = \bra{F_4}T_{4j};
\label{eqn4}
\end{equation}
i.e., take the state $\ket{F}$ and run it backwards through the beam splitters
in Fig.~\ref{fgr1}.  Since TSVF discussions generally use $\bra{\phi_j}$, we present the backward wave in this form:
\begin{align}
\bra{\phi_1} &= \al\bra{A_1}+\bt\bra{Q_1},\quad 
\bra{\phi_2} = \al\bra{A_2} +r\bt(-\bra{B_2} + \bra{C_2}),
\notag\\
\bra{\phi_3} &= \al\bra{A_3} +\bt\bra{E_3},\quad
\bra{\phi_4} = \bra{F_4}.
\label{eqn5}
\end{align}
Here the subscripts are time labels that correspond to those in \eqref{eqn3}.

\xb \outl{TSVF locations: particle only in $A$ at $t_1,t_3$, but in $A$, $B$,
  $C$ at $t_2$; seems odd} \xa

Given the two state vectors, the bra-ket pair ${\bra{\phi_j}~\ket{\psi_j}}$,
Vaidman invokes a phenomenological principle which says that at an intermediate
time $t_j$ the particle can be said to be present in some channel provided the
amplitude for that channel in both $\bra{\phi_j}$ and $\ket{\psi_j}$ is
nonzero; in particular this will produce what Vaidman (Sec.~III of
\cite{Vdmn13}) calls a ``weak trace'' in the channel. Applying this principle
to \eqref{eqn3} and \eqref{eqn5}, one concludes that at $t_3$ the particle was
in $A$: $\ket{A_3}$ is present in $\ket{\psi_3}$ and $\bra{A_3}$ in
$\bra{\phi_3}$. But it was not in $E$: although $\bra{E_3}$ is present in
$\bra{\phi_3}$, $\ket{E_3}$ is absent from $\ket{\psi_3}$. Similarly, at $t_1$
the particle was in $A$ but not in $D$. However, at $t_2$ the particle was
present in both $B$ and $C$ as well as $A$, in disagreement with the claim in
\cite{LAAZ13} that the particle was in $A$ and not elsewhere. Vaidman admits
that his result seems a bit odd, as it is hard to imagine how a particle could
suddenly appear in $B$ and $C$ in the inner MZI when it was not present in
channel $D$ connecting $S$ to the inner MZI and then, equally mysterious,
suddenly disappear so as to be absent from both $E$ and $H$. As we shall see in
Sec.~\ref{sct6}, the TSVF approach interpreted in this way will sometimes give
reasonable results, but can also mislead.

\section{Consistent Histories Analysis}
\label{sct3}

\subsection{Introduction}
\label{sbct3.1}

\xb
\outl{CH a consistent extension of textbook QM; includes microscopic events.
CH references}
\xa

Standard quantum mechanics as found in textbooks is far from a complete theory,
and its rules do not cover situations which precede measurements, such as the
one considered here. The consistent histories (CH) formalism is a systematic
and consistent extension of textbook quantum mechanics. It gives the
same results as the textbooks in the domain where the textbook rules can be
properly applied, but in addition allows a paradox-free discussion of
microscopic properties and events, such as those taking place in the nested
MZI, for which textbooks provide little guidance. The CH approach, also known
as decoherent histories, was developed to some degree independently by various
researchers; some representative references are \cite{Omns99,GMHr14,FrHh14}.
Short introductions will be found in \cite{Hrtl11,Grff14b}, an application to
Bohm's version of Einstein-Podolsky-Rosen in \cite{Grff11b}, and a discussion
of how it relates to various quantum conceptual difficulties in \cite{Grff14}.
A reasonably complete, albeit lengthy, treatment in \cite{Grff02c} provides the
basis for the material that follows.

\xb \outl{CH principles: Qm properties, unitary time
  development, Born probabilities, Qm histories. Measurements not fundamental}
\xa

The basic principles of the CH approach are: quantum properties as defined by
von Neumann, unitary time development using the Schr\"odinger equation,
stochastic dynamics using the Born rule and its extensions, and quantum
histories. Measurements are \emph{not} included in this list, for in the CH
approach measurements (see Sec.~\ref{sct5} below) are simply particular
examples of quantum processes, all of which can analyzed using basic principles
that make no reference to measurements. How these principles permit a detailed
and consistent analysis of the nested MZI ``which path'' problem will be
evident from the following discussion.

\subsection{Quantum properties and sample spaces}
\label{sbct3.2}

\xb \outl{vN: subspace $\lra$ Qm property; Cl property = sset of phase space;
  Cl \NOT, \AND, \OR} \xa

According to von Neumann, Sec.~III.5 of \cite{vNmn32b}, a \emph{property} of a
quantum system at a particular time is represented by a \emph{subspace} of the
quantum Hilbert space, or, equivalently, the projector $P$ (orthogonal
projection operator) onto this subspace.%
\footnote{A finite-dimensional Hilbert space will be adequate for our
  discussion, so the additional requirement that the subspace be closed is not
  needed.} %
Here `property' is used in a restricted sense to mean a physical attribute (a
proposition) which can be true or false; thus `energy' is not a property, but
`the energy is $\ep_0$' or `the energy is less that $\ep_1$' are examples of
properties. The negation of a property, e.g., '`the energy is greater than or
equal $\ep_1$' for the case just mentioned, is also a property.
In classical physics a property in the sense used here always corresponds to a
point or a set of points in the classical phase space, and its negation
to all the points not in this set. Given two properties $P$
and $Q$, the combinations `$P$ \AND\ $Q$' and `$P$ \OR\ $Q$' correspond to the
intersection and union of the two sets of points. If the sets corresponding to
$P$ and $Q$ do not overlap, so the intersection is empty, `$P$ \AND\ $Q$' is
the empty set, the property which is always false, and whose negation, `\NOT\
$P$' \OR\ `\NOT\ $Q$', corresponds to the whole phase space, the property that
is always true.

\xb
\outl{Qm \NOT\ $\lra$ orthog complement; \AND, \OR\ defined if $PQ=QP$ }
\xa

\xb \outl{Four approaches if $PQ\neq QP$: Qm logic, HVs, CH, Ignore problems}
\xa

In quantum mechanics the negation of a property $P$ (following von Neumann) is
the \emph{orthogonal complement} of its subspace, with projector $\tilde P =
I-P$, where $I$ is the identity operator. However, `$P$ \AND\ $Q$' and `$P$
\OR\ $Q$' only have a simple definition in the case in which the projectors
commute, $PQ=QP$, and then $PQ$ is the projector representing `$P$ \AND\ $Q$',
and $P+Q-PQ$ is the projector representing (nonexclusive) `$P$ \OR\ $Q$'. The
headaches of quantum interpretation are all closely linked with the question of
what to do when $P$ and $Q$ do \emph{not} commute. There are various
approaches. The first, quantum logic \cite{BrvN36}, assigns a meaning to `$P$
\AND\ $Q$' using the intersection of the subspaces, whether or not the
projectors commute. However, quantum logic, as the name suggests, necessitates
a change in the rules of reasoning in a fundamental way, and physicists have
not yet had much success in making sense of the quantum world using the new
rules; perhaps our intellectual children or grandchildren will do better. (See
Sec.~4.6 of \cite{Grff02c} for a very simple example showing why new rules are
necessary.) A second approach is to replace or augment the Hilbert space with
\emph{classical} hidden variables which do not suffer from noncommutation
troubles. However, the predictions of hidden variables theories often differ
from those of Hilbert space quantum mechanics, and when these difference are
tested by experiment, the hidden variables approach always loses, so it seems
unlikely that this approach will prove useful in solving quantum mysteries. A
third approach, employed by CH, limits discussions to cases where projectors
commute: the conjunction `$P$ \AND\ $Q$' is only defined if $PQ=QP$, in which
case $PQ$ is the projector that represents the conjunction; but if $PQ\neq QP$,
`$P$ \AND\ $Q$' is undefined, ``meaningless'' in the sense that this
interpretation of quantum mechanics cannot assign it a meaning. Similarly, `$P$
\OR\ $Q$' only makes sense when $PQ=QP$. The CH approach retains the ordinary
rules of reasoning \emph{provided} their application is suitably restricted to
a single domain or ``framework,'' examples of which are given below. A fourth
approach is to go ahead and reason or calculate while ignoring, or at least
paying little attention to, the noncommutation problem. A substantial body of
quantum foundations literature is devoted to discussing the resulting
paradoxes.

\xb \outl{Spin half: $S_z$, $S_x$ projectors don't commute. CH: incompatible
  combinations meaningless} \xa

To make the discussion less abstract, consider the case of a two-state system,
a spin-half particle for which $S_z$, the $z$ component of angular momentum,
can take the values of $+1/2$ or $-1/2$ in units of $\hbar$. Let the projectors
for the corresponding eigenstates be $z^+=\dya{z^+}$ and $z^-$. They commute,
since their product (in either order) is the zero operator, the quantum
property that is always false, and their sum is $I$, the property that is
always true, so they constitute a quantum sample space or \emph{framework},
call it the $Z$ framework. Similarly, the projectors $x^+$ and $x^-$ on the
eigenstates of $S_x$ constitute the $X$ framework. The $X$ and $Z$ frameworks
are \emph{incompatible} in that the projectors in one do not commute with the
projectors in the other, and the CH \emph{single framework} rule prohibits
combining them: `$S_z=+1/2$ \AND\ $S_x=+1/2$' is meaningless, as is `$S_z=+1/2$
\OR\ $S_x=+1/2$'. This is consistent with the fact that $S_z$ can be measured,
and $S_x$ can be measured, but, as the textbooks tell us, they cannot be
measured \emph{simultaneously}. The simple explanation for the textbook rule is
that these combinations of $S_z$ and $S_x$ properties do not correspond to
Hilbert subspaces, so no quantum property exists which could be revealed by a
simultaneous measurement.

\xb
\outl{Qm sample space = PDI. Spin-1/2 example. Compatible PDIs; common
  refinement; single framework rule (SFR)} 
\xa

Standard (Kolmogorov) probability theory makes use of a \emph{sample space} of
mutually-exclusive possibilities, one and only one of which can occur, be true,
in any given situation. A quantum sample space is a \emph{projective
  decomposition of the identity} (PDI), a collection of mutually orthogonal
projectors which sum to the identity. In the simplest situation each of the
projectors has rank one: it projects onto a one-dimensional subspace (`ray') in
the Hilbert space, and the sample space corresponds to an orthogonal basis of
the Hilbert space.  The $X$ and $Z$ frameworks defined above are PDIs and 
thus quantum sample spaces. 
Two sample spaces (PDIs) $\PC$ and $\QC$ are \emph{compatible} if every
projector in one commutes with every projector in the other; otherwise they are
\emph{incompatible} (as in the case of $X$ and $Z$ defined earlier). When
compatible, $\PC$ and $\QC$ possess a \emph{common refinement} made up of all
distinct nonzero products of projectors from $\PC$ with projectors from $\QC$,
which is a PDI and thus a sample space. The single framework rule
requires that logical or probabilistic arguments be based on a \emph{single}
sample space; in particular one cannot use one sample space for part of an
argument and then shift to another, incompatible sample space for another part,
because this leads to paradoxes, such as those discussed in Ch.~22 of
\cite{Grff02c}.

\xb \outl{Qm allows different sample spaces. Choice may not be obvious
  from context. CH regulates this diversity via SFR. Should always know which
  sample space is in use.} \xa

A crucial difference between ordinary (classical) applications of probability
theory and its use in quantum mechanics is that in the latter there are often a
variety of possible sample spaces which can be used to model a physical
process, and whereas in some cases there is an ``obvious'' choice, in other
cases the choice is far from obvious. Sometimes alternative choices of sample
space are useful for different reasons, but there is no way of combining them;
they are incompatible. The CH approach meets this diversity by noting that it
is there, acknowledging that it is sometimes valuable to have two or more
perspectives on a physical problem, and then insists on a strict enforcement of
the single framework rule that prevents this diversity from leading to
paradoxes. Because of the possibility of employing different frameworks it is
important to be clear about which framework is being used in a particular
discussion. Often this is evident from the context, but sometimes it is not,
and then making the choice explicit can be helpful in order to avoid
contradictions and paradoxes.  Further comments on choosing frameworks will be
found in Sec.~\ref{sbct4.2}, but first let us look at some examples. 

\subsection{Histories and consistency}
\label{sbct3.3}

\xb
\outl{Histories sample space for stochastic process. Cl \& Qm example}
\xa

The sample space of a classical stochastic process, such as a random walk or
successive flips of a coin, consists of \emph{ histories}: sequences of
properties at successive times. E.g., flipping a coin three times in a row can
give rise to eight different histories, HHH, HHT, HTH, etc.; H for heads and T
for tails. In quantum mechanics a history consists of a sequence of quantum
properties, thus a sequence of projectors, at successive times.
We will be considering histories for the MZI in Fig.~\ref{fgr1}, and in
particular looking at projectors which in some way identify the location of
the particle in different channels at successive times. Capital letters
$A=\dya{A}$, $B$, etc. denote projectors for these channels, and subscripts
indicate times, as in \eqref{eqn1}. For example, the
history
\begin{equation}
 S_0\od A_2\od F_4 = S_0\od I_1\od A_2\od I_3 \od F_4
\label{eqn6}
\end{equation}
says that the particle started in channel $S$ at $t_0$, was in the $A$ channel
at $t_2$, and in the $F$ channel at $t_4$. The forms on the left and right
sides are equivalent, because the identity operator $I$ provides no information
about the particle at the times $t_1$ and $t_3$. (One can interpret $I_3$ as
either the identity on the full Hilbert space spanned by 15 kets, or simply
$A_3+E_3+H_3$, the possibilities for the particle at $t_3$; for our purposes
these are equivalent.)

\xb
\outl{Symbol $\od$ a form of $\ot$. History Hilbert space. $\FC_A$ is example 
of Qm  history family}
\xa

The $\od$ symbol in \eqref{eqn6} is a variant of $\ot$ and is used to
indicate a tensor product: the history is represented as a tensor product
of projectors on a \emph{history Hilbert space}
\begin{equation}
\breve\HC=\HC\od\HC\od\cdots\HC
\label{eqn7}
\end{equation}
 constructed using copies of the Hilbert space
$\HC$ that describes the system at a single time. A sample space or
\emph{family} of quantum histories is a collection of projectors on $\breve
\HC$ which sum to 
the identity $\breve I$, thus a PDI. For example, the history \eqref{eqn6}
is a member of a family of four histories:
\begin{equation}
  S_0\od A_2\od F_4,\hquad S_0\od \tilde A_2\od F_4,\hquad
S_0\od I_2 \od \tilde F_4,\hquad \tilde S_0 \od I_2 \od I_4.
\label{eqn8}
\end{equation}
A tilde over a letter indicates
negation, thus $\tilde A_2 = I_2 - A_2 = B_2+C_2$, and $\tilde S_0=I_0-S_0$.
Employing the usual rules for adding tensor products of operators, the reader
can easily check that the projectors in \eqref{eqn8} sum to $\breve I =I_0\od
I_2\od I_4$, which means the same thing as $I_0\od I_1\od I_2\od I_3\od I_4$
when all five times are in view. (Once again, replacing each $I_j$ with the
identity on the full Hilbert space of all 15 kets that appear in \eqref{eqn1}
would make no difference in our discussion.)

\xb
\outl{Ignore $\tilde S_0 \od I_2 \od I_4$. Chain kets for $\FC_A$.}
\xa

We shall only be interested in cases in which the particle is in $S$ at $t_0$,
and therefore we shall omit the fourth history in \eqref{eqn8} from the
discussion which follows, which is equivalent to assigning it zero probability.
The three histories that remain can be assigned probabilities using the
\emph{extended Born rule}, which, because they all begin with a pure state
$S_0$, is most easily discussed using \emph{chain kets}, Sec.~11.6 of
\cite{Grff02c}:
\begin{align}
 &\ket{S_0,A_2,F_4} :=  
F_4\, T_{4,2}\, A_2\, T_{2,0}\, \ket{S_0} =\al^2\ket{F_4},
\notag\\
 &\ket{S_0,\tilde A_2,F_4} = 0,\quad 
\ket{S_0, I_2,\tilde F_4} = \bt \ket{H_4} + \al\bt\ket{G_4}.
\label{eqn9}
\end{align}
Here the chain ket $\ket{S_0,A_2,F_4}$ is obtained by applying to $\ket{S_0}$
the sequence of unitary operators and projectors, $T_{2,0}$, $A_2$, $T_{4,2}$,
and $F_4$, the same order as the events in the history (but from right to
left). The other chain kets in \eqref{eqn9} are obtained by the same procedure.
Note that chain kets are elements of the single-time Hilbert space $\HC$, not
the history Hilbert space $\breve\HC$.

\xb \outl{Consistent family $\lra$ orthogonal chain kets. Probability =
  (norm of chain ket)$^2$} \xa

A family of histories is said to be \emph{consistent} if any two chain kets
associated with distinct histories in this family are orthogonal to each other.
For a consistent family the probability assigned to each history by the
extended Born rule---to be precise, the probability conditioned on the initial
state $S_0$---is the square of the norm of its chain ket, the inner product of
the chain ket with itself. The orthogonality just mentioned is referred to as a
\emph{consistency condition}. If it is not satisfied the history family is said
to be \emph{inconsistent}, and no probabilities can be assigned to the
corresponding histories. This means that an inconsistent family cannot be used
in a probabilistic description of a quantum system; it is ``meaningless''
(lacks a meaning) within the CH formulation. (But see the additional comments
in Sec.~\ref{sbct4.3}.)

\xb
\outl{Probabilities, conditional probabilities for $\FC_A$. 
$\Pr(A_2 \vb F_4, S_0) =1$}
\xa

From \eqref{eqn9} it follows that the family in \eqref{eqn8}, with the final
$\tilde S_0$ history omitted, is consistent. The corresponding probabilities
conditioned on $S_0$ are
\begin{equation}
 \Pr(A_2,F_4\vb S_0) = \al^4,\quad \Pr(\tilde A_2,F_4\vb S_0) = 0,\quad 
\Pr(I_2,\tilde F_4\vb S_0) = \bt^2 + \al^2\bt^2,
\label{eqn10}
\end{equation}
and in view of \eqref{eqn2} they sum to 1. It then follows that
\begin{align}
 &\Pr(F_4\vb S_0) = 
\Pr(A_2,F_4\vb S_0) + \Pr(\tilde A_2,F_4\vb S_0)=\al^4,
\notag\\
 &\Pr(A_2,F_4\vb S_0)/\Pr(F_4\vb S_0) = \Pr(A_2 \vb F_4, S_0) =1.
\label{eqn11}
\end{align}
The last equality means that if the particle was in $S$ at $t_0$ and arrived in
$F$ at time $t_4$ it was in channel $A$ at the intermediate time $t_2$.

\subsection{History families using refinements}
\label{sbct3.4}

\xb
\outl{Constructing consistent families using successive refinements}
\xa

The family of three histories
\begin{equation}
  S_0 \od \{F_4, G_4, H_4\}, 
\label{eqn12}
\end{equation}
using a compact notation, involves only two times (or identity operators at the
intermediate times), and is obviously consistent since the chain kets end in
three mutually orthogonal states, so the extended Born rule reduces to the
usual Born rule. A possible strategy for constructing consistent families is to
take each of the histories in \eqref{eqn12} and refine it by replacing $I$ at
some intermediate times with sums of two or more projectors, and then testing
whether the result is consistent. We shall consider refinements of the
subfamily $S_0\od F_4$, but the same techniques can be applied to the other
subfamilies $S_0\od G_4$ and $S_0\od H_4$. Each subfamily can be refined, and
its consistency checked, without regard to refinements of the other
subfamilies; in particular, events at an intermediate time in one subfamily can
be independent of those in a different subfamily. If a refinement yields an
inconsistent (sub)family, further refinement will not restore consistency; one
should try some other possibility.

\xb
\outl{Refinements of $S_0\od F_4$: $\FC_A,\,\FC_B,\,\FC_{ABC}$}
\xa

We have already seen that the family \eqref{eqn8} is consistent, which means
that the (sub)family consisting of the first two of its histories,
\begin{equation}
 \FC_A:\; S_0\od\{A_2,\tilde A_2\}\od F_4 = 
S_0 \od \{A_2,B_2+C_2\}\od F_4,
\label{eqn13}
\end{equation}
a refinement of $S_0\od F_4$, is also consistent. This family can be further
refined by adding events at times $t_1$ and $t_3$ to a family
\begin{equation}
 \FC'_A:\; S_0\od \{A_1,D_1,Q_1\} \od \{A_2,B_2+C_2\} \od
\{A_3,E_3,H_3\} \od F_4,
\label{eqn14}
\end{equation}
of $3\tm 2\tm 3 = 18$ histories. However, the chain kets for all of them
vanish, with the sole exception of the history
\begin{equation}
 S_0\od A_1\od A_2\od A_3\od F_4.
\label{eqn15}
\end{equation}
Consequently, 
\begin{equation}
 \Pr(A_1,A_2,A_3\vb S_0,F_4) = 1,
\label{eqn16}
\end{equation}
which is to say that the particle which entered the nested MZI through $S$ and
left it through $F$ was in the $A$ channel the entire
time it was inside the interferometer.  Also it was not in $D_1$ or $Q_1$ at
time $t_1$, nor was it in $E_3$ or $H_3$ at time $t_3$. The situation at $t_2$
is less clear, and will be discussed further below. 

A different refinement of $S_0\od F_4$ yields the family 
\begin{equation}
 \FC_B:\; S_0\od\{B_2,\tilde B_2\}\od F_4 =
S_0\od\{B_2,A_2+C_2\}\od F_4. 
\label{eqn17}
\end{equation}
It is inconsistent, since the chain kets 
\begin{equation}
 \ket{S_0,B_2,F_4} = -(\bt^2/2)\ket{F_4}, \quad 
 \ket{S_0,\tilde B_2,F_4} = (\al^2+ \bt^2/2)\ket{F_4}, 
\label{eqn18}
\end{equation}
are obviously not orthogonal, at least when $\al$ and $\bt$ are both positive,
as assumed in \eqref{eqn2}.  Hence further refining it, by replacing
$A_2+C_2$ with the pair $\{A_2,B_2\}$, will lead to an inconsistent family
of three histories
\begin{equation}
 \FC_{ABC}:\; S_0\od\{A_2,B_2,C_2\}\od F_4.
\label{eqn19}
\end{equation}

The inconsistent $\FC_{ABC}$ can also be obtained from the consistent $\FC_A$
in \eqref{eqn13} by replacing $\tilde A_2 = B_2+C_2$ with the pair
$\{B_2,C_2\}$. Why should this make a difference? Here we encounter a very
important conceptual difference between quantum and classical physics. If
projectors $B$ and $C$ commute, the quantum counterpart of \OR\ in the sense of
``$B$ or $C$ or both'' is the projector $B+C-BC$, and if, as in the present
instance, $BC=0$, the projector $B+C$. But a Hilbert subspace $B+C$ contains
linear combinations such as $0.8\ket{B}-0.6\ket{C}$ which belong to neither the
$B$ nor the $C$ subspace. In the classical world if something is ``$B$ or
$C$'', assuming $B$ and $C$ are mutually exclusive, we know at once that it is
either $B$ or else it is $C$. In the quantum world this is true \emph{provided}
one is using a framework that contains $B$ and $C$ as separate projectors, but
\emph{not} if one is using the coarser description in which only $B+C$ appears,
not $B$ and $C$ separately. The sample spaces $\{A,B+C\}$ and $\{A,B,C\}$ are
not the same, and it is important to pay attention to which of these is in use.
Some additional discussion of this very important point will be found in
Sec.~\ref{sbct4.1}.

\xb
\outl{$\FC_C$ with $\{C,\tilde C\}_2$ inconsistent unless $\al=\sqrt{1/3}, \bt
  = \sqrt{2/3}$ }
\xa

Yet another refinement of $S_0\od F_4$ is
\begin{equation}
\FC_C:\; S_0\od\{C_2,\tilde C_2\}\od F_4 = S_0\od\{C_2,A_2+B_2\}\od F_4,
\label{eqn20}
\end{equation}
with chain kets
\begin{equation}
 \ket{S_0,C_2,F_4} = (\bt^2/2)\ket{F_4}, \quad 
 \ket{S_0,\tilde C_2,F_4} = (\al^2- \bt^2/2)\ket{F_4}.
\label{eqn21}
\end{equation}
Thus $\FC_C$ will be inconsistent apart from the special case
\begin{equation}
 \al=\sqrt{1/3},\quad \bt = \sqrt{2/3},
\label{eqn22}
\end{equation}
for which the second chain ket in \eqref{eqn21} is zero, allowing one 
to assign probabilities 
\begin{equation}
 \Pr(F_4\vb S_0) = \Pr(C_2,F_4\vb S_0) = \bt^4/4 = 1/9,\quad 
 \Pr(C_2\vb S_0,F_4) = 1,
\label{eqn23}
\end{equation}

\xb
\outl{Inferences $\Pr(A_2|F_4)$, $\Pr(C_2|F_4)$ not contradictory because
derived in incompatible families}
\xa

\xb
\outl{Connection to 3 box paradox, which is discussed in  CQT}
\xa

But does not the result $\Pr(C_2\vb S_0,F_4) = 1$ in \eqref{eqn23}, given the
choice of coefficients in \eqref{eqn22}, contradict the earlier result
$\Pr(A_2\vb S_0,F_4) = 1$ in \eqref{eqn11}? Can a particle emerging in channel
$F$ at time $t_4$ have been with probability 1 in both channel $A$ \emph{and}
in channel $C$ at $t_2$? Is this not a contradiction? No, for in the CH
approach results obtained in two separate frameworks cannot be combined unless
the frameworks themselves can be combined; once again the single framework
rule. All the projectors for histories in $\FC_A$ commute with those in
$\FC_C$, so there is a common refinement, $\FC_{ABC}$ in \eqref{eqn19}. But
this common refinement is an inconsistent family, even for the special choice
of parameters \eqref{eqn22} for which $\FC_C$ is consistent. Hence $\FC_A$ and
$\FC_C$ are incompatible, or \emph{incommensurate} if one wants a separate term
for the situation in which the inability to combine families arises from a
failure of the consistency conditions, and thus the inability to assign
probabilities, rather than the fact that the history projectors do not commute.
The situation just discussed is an instance of the three box paradox of
Aharonov and Vaidman \cite{AhVd91}; see Sec.~22.5 of \cite{Grff02c} for a
discussion of how the CH approach resolves (or ``tames'') this paradox. The
three box paradox has certain features in common with the Bell-Kochen-Specker
paradox \cite{Mrmn93}, some versions of which are considered in Ch.~22 of
\cite{Grff02c}.

\section{Additional remarks}
\label{sct4}

\xb
\outl{CH QM requires different thinking from CP, so some comments}
\xa

The previous discussion has employed some features of stochastic
quantum time development using histories which call for a different type of
thinking than is common in classical physics, and the following comments may
be helpful in indicating how the CH approach avoids paradoxes and
comes to reliable and non-contradictory conclusions about microscopic quantum
events.

\subsection{$B+C$ vs $\{B,C\}$}
\label{sbct4.1}
\xb
\outl{Double slit as analogy for difference between $B+C$ and $\{B,C\}$}
\xa

The distinction between the sum $B+C$ of two projectors and the projectors $B$
and $C$ considered as exclusive properties when $BC=0$ was noted following
\eqref{eqn19}, and can be illustrated using the well-known double slit
experiment. A particle in an initial state $\ket{S}$ travels towards the slit
system and passes through it at an intermediate time before reaching the
interference zone. Let $B$ and $C$ be projectors on two nonoverlapping regions
of space, one containing the upper slit and one the lower slit, such that as it
passes through the slit system the particle wavepacket is in the combined
region corresponding to the projector $J=B+C$. Let $F$ be a projector on
a region of destructive interference. A family
\begin{equation}
 S\od\{J,\tilde J\}\od \{F,\tilde F\}
\label{eqn24}
\end{equation}
with four histories will be consistent, whereas refining it by replacing
$\{J,\tilde J\}$ at the intermediate time with $\{B, C,\tilde J \}$ will result
in an inconsistent family. The projector $J=B+C$ is noncommittal: ``the particle
passed through the slit system, but I tell you no more,'' and is compatible
with later interference, whereas $B$, ``the particle passed through the upper
slit,'' and $C$, ``it passed through the lower slit,'' are not. Feynman in
Ch.~1 of \cite{FyLS65}, with his superb physical intuition, knew that when
discussing interference one should not try and identify which slit the particle
passed through. One can think of the CH rule that excludes inconsistent
families as a mathematical formulation of this intuition, allowing it to be
applied not only to the double slit (for which see Ch.~13 in \cite{Grff02c}),
but to many other situations as well, and used by those of us whose physical
intuition falls somewhat short of Feynman's. Indeed, if one thinks of the $B$
and $C$ arms of the inner MZI in Fig.~\ref{fgr1} as analogous to two slits, it
is easy to understand why specifying them as exclusive alternatives can give
rise to conceptual difficulties, as well as peculiar effects when using weak
measurements, as discussed below in Sec.~\ref{sbct5.2}.

\xb
\outl{$B,C$  $\lra$, ground \& first excited state of harmonic oscillator}
\xa

Yet another example is to think of $B$ and $C$ as projectors on the ground
state and first excited state of a quantum harmonic oscillator. Then both $B$
and $C$ correspond to states of well-defined energies, and ``$B$ or $C$'' could
be taken to mean that the oscillator has one energy or the other. On the other
hand the subspace on which $B+C$ projects includes states which oscillate in
time and do not have a well-defined energy.

\subsection{Multiple frameworks}
\label{sbct4.2}

\xb
\outl{SFR allows \emph{considering}, forbids \emph{combining} incompatible 
frameworks}
\xa

\xb
\outl{Analogy: change of coordinate system. But in Qm case info in $X$ not
  present in $Z$}
\xa

In quantum mechanics, unlike classical physics, there are often several
distinct ways to describe a physical system and its time evolution, each of
which is an acceptable application of quantum principles, but because of
incompatibility they cannot be combined to form a single description. Which
framework to use will be determined by the type of question one wants to
address, and the single framework rule of CH helps guide this choice so as to
achieve reliable results rather than inconsistencies and paradoxes. The single
framework rule does not prohibit constructing multiple frameworks; what it
forbids is \emph{combining} incompatible frameworks to form a single
description. Because there are multiple possibilities, it is important to be
clear about which framework is being used in a particular discussion, something
that may or may not be obvious from the context. Note that the choice of which
framework to use is made by the physicist who is applying quantum principles to
a particular situation; it is \emph{not} determined by some law of nature. In
this respect it is analogous to the choice of a convenient coordinate system in
classical physics with, however, the disanalogy that in classical physics all
the information represented in a particular coordinate system can be transformed
to a different coordinate system in a one-to-one fashion. By contrast, the
information present in, say, the $X$ framework of a spin-half particle is
entirely different from that in the incompatible $Z$ framework.

\xb
\outl{Examples: we assumed $\DC^1\Ra F_4$, but textbooks approach is different}
\xa

\xb
\outl{Alternative refinement of $S_0\od F_4$ using backward wave from $\bra{F}$}
\xa

As an example, with reference to Fig.~\ref{fgr1} we have been assuming, in
agreement with previous literature, that a particle detected in $\DC^1$ was at
time $t_4$ in channel $F$. This is not the only possibility: the framework of
unitary time development that students learn in Quantum 101 employs a projector
$[\psi_4]=\dya{\psi_4}$, where $\ket{\psi_4}=T_{40}\ket{S_0}$ is defined in
\eqref{eqn23}. This unitary framework is a perfectly acceptable quantum
description; there is nothing wrong with it. But it cannot be used to discuss
which channel the particle was in at $t_4$ because $[\psi_4]$ does not commute
with $F_4$, $G_4$, or $H_4$. The CH approach allows an alternative framework in
which at a time just before the measurements take place the particle is in one
of the channels leading to the detectors in Fig.~\ref{fgr1}, and in addition
(see Sec.~\ref{sbct5.1}) it justifies the inference from detection by $\DC^1$
to the particle's having been in $F$ at $t_4$, something which cannot be done
using the textbook approach. So it should come as no surprise that the family
$S_0\od F_4$ can itself be refined in various different ways. For example, the
family
\begin{equation}
 S_0 \od \{[\phi_2], I_2-[\phi_2]\}, F_4,
\label{eqn25}
\end{equation}
where $[\phi_2]=\dya{\phi_2}$ is the projector corresponding to the the
backward wave $\bra{\phi_2}$ defined in \eqref{eqn5}, is a possible refinement
of $S_0\od F_4$; we leave it as an exercise to show that it is consistent. Of
course it is useless for addressing the question of whether the particle is or
is not in the A channel at $t_2$; for that purpose one needs to use $\FC_A$.
Also, as noted in Sec.~\ref{sbct3.4}, both $\FC_A$ and $\FC_C$ are consistent,
but mutually incompatible, families for the choice of parameters in
\eqref{eqn22}; the first is useful for deciding whether the particle was or was
not in $A$, but cannot be used to discuss whether it was in $C$; the second can
address the question of whether or not it was in $C$, but can say nothing about
$A$.

\xb
\outl{Liberty in choosing family $\not\Rightarrow$ contradictions}
\xa

\xb
\outl{Example: $\Pr(E_3)=0$ in $\FC_A$, not defined in $\FC_C$}
\xa

Given this liberty in choosing families, one can ask whether this might
not give rise to contradictions: different families assigning different
probabilities to some event at an intermediate time, say $A_2$. However, as
long as probabilities are conditioned on the same set of events, e.g., $S_0$
and $F_4$, the (conditional) probability for an event at an intermediate time
will be independent of the consistent family to which it belongs; see the
discussion in Ch.~16 of \cite{Grff02c}. For example, given $S_0$ and $F_4$, the
probability is zero that the particle was in $E$ at time $t_3$. This can be
shown using the family $S_0\od \{E_3,\tilde E_3\}\od F_4$ (or by calculating
the weak value of $E_3$ at time $t_3$ using the method indicated in
Sec.~\ref{sct6}), and the answer is the same if $\FC_A$ is refined by replacing
$I_3$ with $\{E_3,\tilde E_3\}$.
But if $E$ was empty at time $t_3$, how is it possible (see Fig.~\ref{fgr1})
for a particle arriving in $F$ at $t_4$ to get there from $C$ at $t_2$, as must
have been the case according to family $\FC_C$? The answer is that refining
$\FC_C$ by replacing $I_3$ with $\{E_3,\tilde E_3\}$ makes it inconsistent, and
thus when using $\FC_C$ it is meaningless to ask whether the particle was
in channel $E$ at time $t_3$. Once again it is the single framework rule, whose
central role in CH cannot be overemphasized, that prevents combining
incompatible families to arrive at a contradiction. The quantum world is indeed
weird from the point of view of classical physics, which is all the more reason
why it must be analyzed using conceptual and mathematical tools that do not
lead to contradictions and unresolved paradoxes.

\xb \outl{Alpha particle emitted in S wave moving on straight line to detector}
\xa

As an example of multiple incompatible frameworks in a different context,
consider an experiment in which a nucleus decays by emitting an alpha particle
in an S wave (spherical symmetry), which is then detected some distance away.
The experimenter will think of the particle as traveling along an almost
straight path from the source to the detector, and the projectors appropriate
to this description, corresponding to wave packets with a relatively narrow
angular spread, do not commute with those that represent a spherical wave, so
the two descriptions cannot be combined. In the CH approach both the spherical
wave and the narrow wave packet constitute perfectly acceptable quantum
descriptions, and one or the other may be more useful for certain purposes.

\xb
\outl{Cl intuition fails because \emph{unicity} not valid in QM}
\xa

The notion of multiple possible descriptions of the same experiment, of a sort
that cannot be combined with each other, is very different from what one
encounters in classical physics, so it may be helpful to try and identify the
point at which classical intuition fails. In the world of everyday experience,
where a classical approximation to quantum theory is adequate for all practical
purposes, we tend to believe that at any instant of time there is a unique
state of the world that is true or actual or real, even though no one knows
what it is. This belief, elsewhere referred to as \emph{unicity} (Sec.~27.3 of
\cite{Grff02c}), has a mathematical counterpart in the phase space of classical
mechanics, where the state of a mechanical system at a given time is
represented by one and only one point in the classical phase space. All
properties (collections of points) that contain this point are true, while
those that do not contain it are false.
A Hilbert space is somewhat analogous to a classical phase space, and its
one-dimensional subspaces, or rays, are analogs of the individual points in the
phase space. But unlike two distinct points in the phase space, two different
rays do not represent mutually exclusive physical properties unless they are
orthogonal to each other. To put the matter differently, if one thinks of a
single ray as representing the ``real'' state of the quantum world, and that
all subspaces that contain it are true, while those orthogonal to it are false,
this leaves many subspaces that belong to neither category, and thus are
neither true nor false. Hence if the real world is best described using a
quantum Hilbert space and its subspaces, rather than a classical phase space,
unicity does not correspond to physical reality.

\subsection{Dynamics and consistency}
\label{sbct4.3}

\xb
\outl{Consistency depends on unitary dynamics as well as projectors}
\xa

\xb
\outl{Nested MZ examples: change $\al,\bt$; remove beamsplitters 3 \& 4}
\xa

It is worth noting that consistency depends not just on the history projectors,
but also on the unitary dynamics, the $T_{kj}$, used to compute the chain kets.
A family which is inconsistent for a particular unitary dynamics may be
consistent for a different dynamics. Thus $\FC_C$ in \eqref{eqn20} is in
general inconsistent, but for the special choice of $\al$ and $\bt$ in
\eqref{eqn22} it is consistent. 
A more drastic change in the dynamics would be to eliminate beam splitters 3
and 4 in Fig.~\ref{fgr1}, in which case the family
\begin{equation}
S_0\od\{A_2,B_2,C_2\}\od\{F_4,G_4,H_4\}
\label{eqn26}
\end{equation}
will be consistent, in contrast to
the inconsistent family $\FC_{ABC}$ in \eqref{eqn19}. Since the particle only
encounters beam splitters 3 and 4 \emph{after} $t_2$, one might be tempted
to suppose that the future is somehow influencing the past. But the change
is in what can inferred about past properties, the particle's location at
$t_2$, from later measurement outcomes, and it is not unreasonable to suppose
that altering the unitary time evolution connecting the two will make a
difference.

\xb
\outl{Interaction with environment--decoherence, external measurement--can
  change consistency}
\xa

There are many other examples. An inconsistent family for an isolated system
may become consistent if that system interacts with an environment. Decoherence
can have this effect, and so can subjecting a system to external measurements.
In the CH approach measurements must themselves be described, at least in
principle, using quantum mechanics, so what can be consistently said about a
system in the presence of a measurement may or may not be possible when there
is no measurement. See the discussion in the paragraph following \eqref{eqn35}
in Sec.~\ref{sbct5.2} for a particular example.

\section{Measurements}
\label{sct5}

\subsection{Introduction}
\label{sbct5.1}

\xb \outl{Measurement: micro info $\ra$ pointer. SQM:
  Born rule $\ra$ probability ``if measured''}

\xb
\outl{CH description of measurement. Pointer PDI, + retrodiction
to microscopic PDI}
\xa

A quantum measurement is a process by which information about some microscopic
property or behavior of the system of interest is amplified so that it can be
represented through distinctive macroscopic properties of a measuring device,
``pointer positions'' in the archaic but picturesque language of quantum
foundations. In textbook quantum mechanics students learn how to calculate a
probability for a microscopic property, such as $S_z=+1/2$ for a spin half
particle, by using the Born rule applied to a ket or density operator for the
microscopic system, and are told that this is the probability of this property
\emph{if it is measured}. The CH approach, see Chs.~17 and 18 of
\cite{Grff02c}, supplies the steps missing from textbooks by providing a
complete, albeit schematic, quantum mechanical description of the entire
measurement process, assuming an appropriate interaction between the apparatus
and the system to be measured. The infamous measurement problem of quantum
foundations, the fact that unitary time development will typically leave the
apparatus in a superposition of pointer states, is disposed of by using a
framework of macroscopic properties, an appropriate PDI corresponding to
different pointer positions. The second measurement problem, inferring the
prior microscopic state from the final pointer position, is taken care of by
using a framework that includes an appropriate microscopic PDI at a time just
before the measurement takes place, and then using standard probabilistic
reasoning to infer (retrodict) the earlier microscopic state from the later
pointer position. In the case of the nested MZI in Fig.~\ref{fgr1} one can
think of $\DC^1$, $\DC^2$ and $\DC^3$ as constituting a single measurement
device whose ``pointer'' is whichever device has detected the particle, while
the microscopic PDI consists of $\{F_4,G_4,H_4\}$, the possible locations of
the particle just before detection. This is how the CH approach justifies the
inference from detection by $\DC^1$ to the particle having been in $F$ at
$t_4$.

\subsection{Weak measurements}
\label{sbct5.2}

\xb \outl{Weak measurement defined. Distinct from weak value, which is not used
  here} \xa

A \emph{weak measurement} in contrast to a \emph{strong} or \emph{projective}
measurement of the type discussed above, is one in which the system to be
measured (in our case the particle or photon) interacts weakly with the
measuring apparatus, so that on average neither the apparatus nor the particle
is strongly perturbed. Hence extracting useful information requires
repeating the experiment a large number of times. (We are not considering the
case in which a large number of weak measurements are carried out in succession
on a single system.) Even though the interaction is weak it can still on rare
occasions produce a strong effect on the measured system; see, for example,
Feynman's discussion in Sec.~1-6 of \cite{FyLS65}.
Though outcomes of weak measurements are often analyzed in terms of \emph{weak
  values}, as in \cite{LAAZ13}, this is not necessary. The mathematical
definition (see \eqref{eqn36} below for an example) of a weak value is clear,
but its physical significance is obscure, and therefore we shall make no use of
it, but instead employ a more straightforward interpretation of the weak
measurement outcome.

\xb
\outl{Qubit probe for each channel in nested MZI. Special $B+C$ probe.
  Interaction defined.} 
\xa

To study the passage of the particle through the nested MZI, assume that
attached to each channel is a two-state system, a qubit probe, which is
initially in its ``ground'' state $\ket{0}$. The probes in channels  $A$, $D$,
$B$, $C$, $E$ are labeled by the corresponding lower case letters $a$, $d$,
$b$, $c$, $e$. In addition there is a special probe $w$ to detect a
particle passing through $B+C$ without distinguishing $B$ from $C$;
recall the discussion in Sec.~\ref{sbct4.1}.
No probes are needed for channels $F$, $G$, and $H$, as these
terminate in strong measurements. The passage of a particle through channel
$P$ with probe $p$ results in a unitary time development
\begin{equation}
 \ket{P}\ot\ket{0}_p \ra \ket{P} \ot 
\bigl(\zt\ket{0}_p + \eta\ket{1}_p\bigr);\quad
\eta = \sqrt{\ep}\,,\;\zt=\sqrt{1-\ep},
\label{eqn27}
\end{equation}
where $\ep$ is a very small number, think of $1/10000$, whereas if the particle
does not pass through the $P$ channel the probe remains in the state
$\ket{0}_p$. For interaction with the $B+C$ probe $w$, use \eqref{eqn27}
twice, once with $P=B$ and once with $P=C$, with $p=w$ in both cases.
It will be convenient to label states of the entire system of probes using a
symbol $\kp$, where $\kp=o$ is the initial state with no probes excited,
$\kp=db$ means probes $d$ and $b$ are excited and the rest are not, and so
forth. Thus when the particle passes through channel $P$ the result is
\begin{equation}
 \ket{P} \ot \ket{\kp} \ra 
\ket{P} \ot \bigl( \zt\ket{\kp} + \eta\ket{\kp p}\bigr),
\label{eqn28}
\end{equation}
where $\kp p$ means $p$ if $\kp=o$, $bp$ if $\kp = b$, and so forth.

\xb
\outl{Probes measured after run: $\ket{1} \Ra$ particle present;
$\ket{0}$: no info}
\xa

\xb
\outl{Final probe measurement can't affect earlier particle trajectory}
\xa

\xb \outl{Info about trajectory from which probes excited if particle emerges
  in $F$ or $G$ or $H$ } \xa

After a given run is finished each probe can itself can be subjected to a
strong measurement in the $\ket{0}, \ket{1}$ basis to determine its value. A
probe state $\ket{1}$ indicates that the particle was in that channel (or in
$B+C$ for probe $w$), but if the state is $\ket{0}$ one learns nothing: the
particle might have been in the channel, but if so it left no trace. Note that
the process of measuring the probes, which takes place after the particle has
completed its path through the trajectory, has no effect upon that trajectory,
since the future does not influence the past; instead, the measurement yields
information about the state of affairs at the earlier time.
One can then ask: given that the particle emerged in $F$ or $G$ or $H$ (as
indicated by its triggering $\DC^1$ or $\DC^2$ or $\DC^3$), which, if any, of
the probes registered its passage through one of the preceding channels? Since
$\ep$ is very small, the answer will usually be ``none at all,'' but
occasionally one of the probes will be excited, and much less frequently two,
or even three probes will have been excited in the very same run, hence
providing information on the trajectory of a single particle during that run.

\xb
\outl{Case where $B+C$ probe present, but $B$, $C$ probes absent}
\xa

\xb
\outl{State $\ket{\Psi_4}$ of particle + probes at $t_4$; its interpretation}
\xa

The discussion is simplest for the case in which the $B$ and $C$ probes are
absent, but the $B+C$ probe is present, along with the probes for $A$, $D$, and
$E$. Let $\ket{\Psi_j}$, the counterpart of $\ket{\psi_j}$ in \eqref{eqn3}, be
the result of unitary time evolution of the particle together with the system
of probes up to time $t_j$, starting from the state $\ket{\Psi_0} =
\ket{S_0}\ot\ket{o}$, and assuming that at time $t_j$ the interaction with the
corresponding probe has just taken place. All the information of interest to us
will be present at time $t_4$, and it is convenient to write $\ket{\Psi_4}$ in
the form
\begin{equation}
 \ket{\Psi_4} = \sum_\kp \ket{\Phi^\kp} \ot \ket{\kp}.
\label{eqn29}
\end{equation}
A straightforward calculation yields
\begin{align}
 \ket{\Phi^o} &= \zt\al\ket{\bar A_4} + \zt^2\bt \ket{H_4},\quad
\ket{\Phi^a} = \eta\al \ket{\bar A_4},
\notag\\
\ket{\Phi^d} &= \ket{\Phi^w} =\zt\eta\bt\ket{H_4},\quad
\ket{\Phi^{dw}} = \eta^2\bt\ket{H_4},
 \label{eqn30}
\end{align}
and all the other $\ket{\Phi^\kp}$, such as $\ket{\Phi^{ad}}$ are zero. We have
used the abbreviation
\begin{equation}
 \ket{\bar A_4} = \al\ket{F_4} + \bt\ket{G_4}  = T_{43}\ket {A_3}
\label{eqn31}
\end{equation}
for the state at $t_4$ which results when $\ket{A_3}$ passes through 
the final beam splitter. One can use these results to derive
probabilities conditioned on the initial state $\ket{\Psi_0}$, such as
\begin{align}
 &\Pr(F_4,a) = |\inpd{F}{\Phi^a}|^2 = \ep\al^4,
 \notag\\
 &\Pr(F_4,o) = |\inpd{F}{\Phi^o}|^2 = (1-\ep) \al^4,
 \notag\\
 &\Pr(F_4) = \al^4,\quad \Pr(a\vb F_4) = \ep.
\label{eqn32}
\end{align}
That is, given that the particle emerged in $F$ at $t_4$ (was detected by
$\DC^1$), there is a conditional probability of $1-\ep$ that no probes were
triggered, $\ep$ that the $a$ probe was triggered, and zero that any other
probe was triggered. In particular, the $d$, $w$, and $e$ probes were never
triggered if the particle emerged in $F$, indicating that this particle was
never in the $D$ or the $E$ channel, and never in the $B+C$ channel system. The
conclusion is the same if the particle emerged in $G$. All of this is
consistent with the discussion of particle trajectories in Sec.~\ref{sbct3.3}.
And it agrees with the conclusion reached by Li et al.\ \cite{LHZZ15}, who
suggested a possible, albeit rather difficult, way to realize the $w$ probe in
an actual experiment. If, on the other hand, the particle emerged in $H$, there
is a probability of order $\ep$ that either the $d$ or the $w$ probe was
triggered, and a probability of order $\ep^2$ that both probes were triggered
in the same run. The $e$ probe is never triggered. Again, this is just what one
might expect.

\xb
\outl{Case where $B$, $C$ probes present, $B+C$ probe absent. 
$\ket{\Psi_4} = \cdots$}
\xa

Next consider the situation in which the $B$ and $C$ probes are present, but
the $B+C$ probe $w$ is absent. A straightforward but somewhat tedious
calculation shows that the nonzero $\ket{\Phi^\kp}$ in \eqref{eqn29} are:
\begin{align}
 &\ket{\Phi^o} = \zt\al\ket{\bar A_4} + \zt^2\bt\ket{H_4},\quad
 \ket{\Phi^a} = \eta\al\ket{\bar A_4},\quad
 \ket{\Phi^d} = \zt\eta\bt\ket{H_4},
\notag\\
 &\ket{\Phi^b} = \hf\zt\eta\bt(-\zt\ket{\bar E_4}+ \ket{H_4}),\quad
\ket{\Phi^{db}} = (\eta/\zt) \ket{\Phi^b},
\notag\\
 &\ket{\Phi^c} = \hf\zt\eta\bt(\zt\ket{\bar E_4}+ \ket{H_4}),\quad
\ket{\Phi^{dc}} = (\eta/\zt) \ket{\Phi^c},
\notag\\
 &\ket{\Phi^{be}} = -\ket{\Phi^{ce}} = -\hf\zt\eta^2\bt\ket{\bar E_4},\quad
 \ket{\Phi^{dbe}}= -\ket{\Phi^{dce}} = -(\eta/\zt) \ket{\Phi^{be}},
\label{eqn33}
\end{align}
where $\ket{\bar A_4}$ is defined in \eqref{eqn31}, and
\begin{equation}
\ket{\bar E_4} := \bt\ket{F_4} - \al\ket{G_4} = T_{43} \ket{E_3},
\label{eqn34}
\end{equation}
is the state produced when $\ket{E_3}$ pass through beam splitter 4.

\xb
\outl{Lists of triggered probes for particle emerging in $F$ or $G$ or $H$}
\xa

Using these results one can determine which probes have been triggered and with
what probability if the particle emerges in one of the channels $F$, $G$, or
$H$. For our purposes the essence of the matter can be summarized in two lists:
the first indicates which probes  can have been excited if the particle emerges in $H$ (detected
by $\DC^3$); and the second gives this information if the particle emerges in
either $F$ or $G$ (detected by $\DC^1$ or $\DC^2$):
\begin{align}
 \text{$H$: }\; &o,\;d,\; b,\; c, \; db, \; dc, \notag\\
 \text{ $F$ \OR\ $G$: }\; &o,\; a,\; b,\; c, \; 
db,\; dc,\;be,\; ce,\;dbe,\; dce. 
\label{eqn35}
\end{align}
Assuming neither $\al$ nor $\bt$ is very small, the probability that a set
$\kp$ of probes was excited is of order $\ep^{|\kp|}$: 1 if no probes have been
excited; and $\ep$, $\ep^2$, or $\ep^3$ in the case of one, two, or three
probes excited during the same run.

\xb
\outl{Interpretation of lists. Pertubing effects of weak measurements in $B$,
  $C$}
\xa

The $H$ list in \eqref{eqn35}, which does not contain $a$ or $e$, is consistent
with the idea that when detected by $\DC^3$ the particle was earlier in the
upper arm of the nested MZI and never in either $A$ or $E$. This is not
surprising. In runs in which the particle was detected by $\DC^1$ or $\DC^2$,
so emerged from the MZI in $F$ or $G$, a single probe $a$ or $b$ or $c$ was
excited with a probability of order $\ep$, but never $d$ or $e$, a result which
could be taken to support Vaidman's assertion, Sec.~\ref{sbct2.2}, that this
particle was in $B$ or $C$ as well as in $A$, but was never in $D$ or $E$.
However, the coincidences, two or more probes triggered during a single run,
agree with the alternative explanation given in \cite{LAAZ13}: the perturbing
effects of a weak measurement in $B$ or in $C$. Thus if the $b$ probe was
excited, it indicates that the particle was in the $B$ channel, not the $C$
channel (note that $b$ and $c$ never appear in coincidence). This spoils the
coherence between the $B$ and $C$ channels, and allows the particle to emerge
from the inner MZI with equal probability in $E$ or in $H$. If it emerges in
$E$ there is a small probability (another factor of $\ep$) that it will trigger
the $e$ probe before reaching either $F$ or $G$. This explains the $be$
coincidences, and the fact that $e$ is never excited unless preceded by $b$ or
$c$. The same reasoning can explain the $db$, $dbe$, $ce$, $dc$, and $dce$
coincidences. In the limit $\ep\ra 0$ this symmetry-breaking effect of the $b$
and $c$ probes will go to zero, and the situation will resemble the one
discussed previously in which these probes were absent and only the $w$ probe
was present. Hence the weak measuring results are consistent with the
conclusions in Sec.~\ref{sbct3.3} based on the CH analysis, where there were no
weak measurements, once one has taken into account the fact that measurements,
even when they are weak, can sometimes perturb a quantum system.

\xb
\outl{$B$, $C$ and also $B+C$ probes all present: results consistent with 
previous  analysis}
\xa

The situation in which the $B$, $C$ and $B+C$ probes are present along with
those for $A$, $D$, and $E$ leads to longer and messier expressions, since
there are many more $\kp$ for which $\Phi^\kp$ is nonzero. However, the results
are consistent with what one would expect from the preceding analysis. In cases
in which $b$ or $c$ are excited, $w$ can also appear (with a probability
smaller by order $\ep$), but if $w$ is not accompanied by $b$ or by $c$ in the
same run, it also is not accompanied by $e$, i.e., the particle always emerges
from the inner MZI in channel $H$.

\xb
\outl{Gaussian probes will not give different results}
\xa

The reader might wonder whether replacing the qubit probes employed here with
Gaussian probes of the sort often employed in the weak measurement literature
would lead to different conclusions. The answer is that it would not. The
easiest way to see this is to note that the interaction specified by
\eqref{eqn27} and \eqref{eqn28} gives rise, so far as the particle (photon) is
concerned, to a noisy quantum ``phase damping'' or``phase flip'' channel (see,
e.g., Sec.~8.3.6 of \cite{NlCh00}), whereas the probe forms the
\emph{complementary channel}, as defined, for example, in \cite{HlGv12}. Since
the phase damping channel has only two Kraus operators, the simplest
complementary channel is two dimensional, thus a qubit channel. A standard
result in quantum information theory is that the direct (phase flip) channel
determines a unique complementary (probe) channel up to a unitary
transformation on the latter \cite{HlGv12}. Thus a Gaussian probe cannot carry
away more information than a qubit probe, though analyzing the Gaussian
probe might be less straightforward.

\section{Two State Vector Formalism}
\label{sct6}

\xb
\outl{Weak value $\avg{}_w$ formula relates chain kets to bra-ket pairs of
  TSVF. Weak values of $A_2,\,B_2,\,C_2$}
\xa

The connection between the two state vector formalism (TSVF)
\cite{AhVd90,Vdmn10} and the CH approach can be conveniently discussed using
the formula
\begin{equation}
 \avg{Q}_w = \mted{\phi_2}{Q}{\psi_2}/\inpd{\phi_2}{\psi_2}
\label{eqn36}
\end{equation}
which defines the \emph{weak value} \cite{AhAV88} of the operator $Q$
in terms of bra-ket pair $\bra{\phi_2}~\ket{\psi_2}$  
at the time $t_2$.  
In particular 
\begin{equation}
 \inpd{F_4}{S_0,P,F_4} = \inpd{\phi_2}{\psi_2}\avg{P}_w
\label{eqn37}
\end{equation}
relates the chain ket, see \eqref{eqn9}, for the history $S_0\od P\od F_4$ to
the weak value of the projector $P$. Using $\ket{\psi_2}$ from \eqref{eqn3} and
$\bra{\phi_2}$ from \eqref{eqn5} one obtains:
\begin{equation}
 \avg{A_2}_w =1,\quad \avg{B_2}_w =-\bt^2/2\al^2, \quad 
\avg{C_2}_w = \bt^2/2\al^2.
\label{eqn38}
\end{equation}

\xb
\outl{$P$ projector at $t_2$: consistency of $S_0\od\{P,\tilde P\}\od F_4$
iff $\avg{P}_w =0$ or 1}
\xa

Since $\avg{}_w$ is linear and $\avg{I}_w = 1$, it is the case that
\begin{equation}
 \avg{P}_w + \avg{\tilde P}_w =1,
\label{eqn39}
\end{equation}
with $\tilde P = I-P$. Consequently, the family $S_0\od\{P,\tilde P\}\od F_4$
will be consistent---one of the chain kets, see \eqref{eqn37}, is zero---if
$\avg{P}_w$ is 1 or 0, but will be inconsistent in all other cases. Thus an
immediate consequence of \eqref{eqn38} is that the family $\FC_A$ with $P=A_2$,
see \eqref{eqn13}, is consistent for all values of $\al$ and $\bt$; $\FC_B$
with $P=B_2$, see \eqref{eqn17}, is never consistent for $\al$ and $\bt$
satisfying \eqref{eqn2}; and $\FC_C$ with $P=C_2$, see \eqref{eqn20}, is only
consistent when $\bt^2/2\al^2=1$, i.e., for the special values in
\eqref{eqn22}.

\xb
\outl{Compare with Vaidman's approach in Sec.~\ref{sbct2.2}}
\xa

\xb
\outl{Histories with more intermediate times: no obvious connection with TSVF}
\xa

Vaidman's principle, as noted in Sec.~\ref{sbct2.2}, is that the particle is
present (in some sense) whenever the weak value of the projector representing
the channel is nonzero, and absent when the weak value is 0. The CH approach
says the particle is present when the weak value of the channel projector is 1,
is absent when the weak value is 0, and otherwise its presence or absence
cannot be discussed, since the history family is inconsistent, so one cannot
assign probabilities. The same comparison can be made if the intermediate time
is $t_1$ or $t_3$, using the bra-ket pair for this time, and assuming a family
of histories defined at $t_0$, $t_4$, and with only one nontrivial (the event
is not simply $I$) intermediate time. Consistency conditions (orthogonality of
chain kets) can also be discussed for histories with additional nontrivial
intermediate times, but for these there is no obvious connection with the TSVF.

\section{Counterfactual Communication}
\label{sct7}

\xb \outl{Problems with Salih et al.\ (Sao) ctfl communication claim} \xa

While the preceding analysis disagrees with Vaidman's claims about the path of
a particle in a nested MZI, it also casts serious doubt upon the counterfactual
communication claim of Salih et al.\ \cite{Slao13}, and, indeed, for much the
same reason: the impossibility of including $B$ and $C$ separately, rather than
$B+C$, at time $t_2$ in a consistent family of histories. The nature of the
difficulty is most easily seen in the reply of Salih et al.\ \cite{Slao14} to
Vaidman's criticism \cite{Vdmn14} of their earlier work in \cite{Slao13}. This
reply contains a figure similar to our Fig.~\ref{fgr1}, albeit rotated by
$45\dg$, and uses identical labels for channels $A$, $B$, $C$, $D$, and $E$,
and similar labeling for the detectors, apart from
subscripts in place of our superscripts.
With reference to this figure Salih et al.\ \cite{Slao14} say that:
\begin{quote} 
  A click at $D_1$ implies that the photon should have followed
  path $A$, and the probability of its existence in the public channel is zero.
\end{quote} 

\xb
\outl{``Public channel'' $\lra$ $C$. Sao claim that $C$ is empty analyzed using
propositions P1, P2, P3.}
\xa

The ``public channel'' in the counterfactual communication protocol is the one
by which Bob communicates with Alice. In terms of Fig.~\ref{fgr1}, all the
beam splitters lie in Alice's domain, and only the $C$ channel mirror belongs to
Bob. Thus for our purposes the public channel is the same as the $C$ channel.
Let us assume in addition that detection by $D_1$, i.e., $\DC^1$, is equivalent
to the particle emerging from the MZI in channel $F$, and consider three
propositions expressed in the notation of Fig.~\ref{fgr1}:
\renewcommand\descriptionlabel[1]{\hspace{\labelsep}\textnormal{#1}}
\begin{description}
 \item[P1.] The particle was in $S$ at $t_0$ and in $F$ at $t_4$.
 \item[P2.] The particle was in $A$ at $t_2$
 \item[P3.] The particle was not in $C$ at $t_2$.
\end{description}
The quotation from \cite{Slao14} given above can be summarized as: P1 implies
P2, P2 implies P3, and therefore P1 implies P3.

\xb
\outl{Sao: P1 $\Ra$ P2 $\Ra$ P3. Step P2 $\Ra$ P3 not correct. Intermediate P2'}
\xa

Let us now examine this argument. The step from P1 to P2 can be justified using
the family $\FC_A$, \eqref{eqn8}, since the final equality in \eqref{eqn11} is
$\Pr(A_2\vb S_0,F_4)=1$. The trouble is with the step from P2 to P3. To
understand why, it is helpful to insert between P2 and P3 the proposition
\begin{description}
 \item[P2\'{}.] The particle was not in $B+C$ at $t_2$.
\end{description}
Since $B_2+C_2=\tilde A_2$ is in $\FC_A$ and ${\Pr(B_2+C_2\vb S_0,F_4)}=0$,
P2\'{} is a direct consequence of P1 as well as implied by P2. 
However, to get from P2\'{} to P3 it is necessary to go from
``not $B+C$'' to ``not $C$'', and this requires refining the framework
containing the projector $B+C$ to one containing both $B$ and $C$. This
nontrivial requirement was noted at the end of Sec.~\ref{sbct3.3}, and
discussed further in Sec.~\ref{sbct4.1}. In the present context such a
refinement would lead to the inconsistent family $\FC_{ABC}$ in \eqref{eqn19},
so it is not allowed.

\xb
\outl{Cannot use P2 $\Ra$ P3 while forgetting P1 $\Ra$ P2 framework, unlike 
Cl physics}
\xa

Note that if one were only concerned about events at time $t_2$ the step from
P2 or P2\'{} to P3 would cause no difficulty; one would simply refine $\{A_2,
B_2+C_2\}$ to $\{A_2,B_2,C_2\}$ and employ the latter to reason from the
presence of the particle in $A$ at time $t_2$ to its absence from $C$. The
difficulty arises because one wants to infer P3 from P1, and P1 contains
information about events at $t_0$ and $t_4$. The single framework rule says
that cannot simply forget the framework used to infer P2 (or P2\'{}) from P1
when carrying out the next step from P2 (or P2\'{}) to P3. It is at this point
where classical reasoning is inadequate. The single framework rule is not part
of the logic of classical physics because it is never needed: all of classical
physics, as seen from a quantum perspective, requires only a single framework.
But in the quantum world one has to modify classical reasoning if one is to
reach reliable conclusions.

\xb
\outl{Direct P1 $\Ra$ P3 needs inconsistent $\FC_C$. Special $\al,\bt$ $\ra$ 
consistent $\FC_C$ $\Ra$ P1 $\Ra$ NOT P3}
\xa

Could one get from P1 to P3 by a direct route that does not include P2? The
coarsest framework that includes $C_2$ along with $S_0$ and $F_4$, and hence
both the premises in P1 and the consequences in P3, is $\FC_C$, \eqref{eqn20},
and in general this family is inconsistent, so it cannot be used to assign a
meaningful probability to $C$ at $t_2$. Only for the special choice of
parameters in \eqref{eqn22} is $\FC_C$ consistent, and in that case one can use
$\FC_C$ to calculate $\Pr(C_2\vb S_0,F_4)$. But this probability is equal to 1,
see \eqref{eqn23}, not 0. Thus for the parameters in \eqref{eqn22} P1 implies
not that P3 is true, but that it is false! (As noted at the end of
Sec.~\ref{sbct3.3}, this result obtained using $\FC_C$ does not contradict
$\Pr(A_2\vb S_0,F_4)=1$ obtained using the family $\FC_A$, which is valid in
general, including the choice of parameters in \eqref{eqn22}, because $\FC_A$
and $\FC_C$ are incompatible---to be precise, incommensurate---families, and
the single framework rule means they cannot be combined.) 

\outl{Salih used simplified example. But if it's wrong why take the rest
  seriously?}
\xa

Thus the inference from P1 to P3 does not satisfy the rules for quantum
reasoning, and one cannot conclude that a particle emerging in channel $F$ was
earlier absent from channel $C$. Hence the argument employed by Salih et al.\
in \cite{Slao14} is not valid. To be sure, the figure in \cite{Slao14} was
presented as a simplified example to illustrate the point the authors were
trying to make; their full protocol is much more complicated. But if the
reasoning applied to this simplified example is defective in the manner just
discussed, it is hard to accept their claim about the more complicated protocol
unless and until it has been justified by better arguments than have been
presented up to now.

\xb
\outl{Further difficulty: Info transmission without interaction is impossible}
\xa

One may add that the very notion of counterfactual communication seems
problematical in light of the fact that it is impossible to transmit
information between quantum systems which do not interact with each other
\cite{Grff11}. Of course, ``interaction'' is not the same thing as sending
particles, though it is hard to see how in the protocol under consideration
there could be an interaction sufficient to convey information in the complete
absence of particles (photons) passing from Bob to Alice. In addition, the
claim in \cite{Slao13} is not that precisely zero particles are involved, but
rather that the number in the Bob to Alice channel can be arbitrarily small in
an asymptotic limit of a large number of opportunities for the photon to pass
back and forth. But then a proper analysis of the situation requires
appropriate quantitative estimates based on sound quantum principles.

\section{Conclusion}
\label{sct8}

\xb
\outl{Particle location at $t_2$. Vaidman's incorrect; Li et al.\ correct}
\xa

The possible paths followed by a particle (photon) that enters the nested
Mach-Zehnder interferometer in Fig.~\ref{fgr1} through channel $S$ and later
emerges in channel $F$ to be detected by $\DC^1$ have been analyzed using
consistent histories. The consistent family $\FC_A$ in \eqref{eqn13} and its
refinement in \eqref{eqn14} leads to the conclusion \eqref{eqn16} that the
particle was in the $A$ arm of the interferometer at all times while inside the
interferometer, and was not in the small interferometer in the sense that zero
probability is assigned to the projector $B+C$ at time $t_2$. This result
agrees with Li et al.\ \cite{LAAZ13} rather than Vaidman \cite{Vdmn13}.

\xb
\outl{However, 0 probability for $B+C$ $\not\Rightarrow$ $B$, $C$ each have 0 
probability}
\xa

However, closer inspection shows that this result is not altogether
straightforward; one needs to pay attention to certain subtleties. Assigning
zero probability to $B+C$ at time $t_2$ conditional on $S_0$ and $F_4$ does not
by itself mean that zero probability can be assigned to $B$ and $C$ separately.
Whereas $B+C$ at time $t_2$ is part of a consistent family $\FC_A$,
\eqref{eqn13} (and $\FC'_A$, \eqref{eqn14}), refining $\FC_A$ by replacing the
projector $B+C$ with the pair $\{B,C\}$, i.e., treating $B$ and $C$ as mutually
exclusive alternatives, leads to an inconsistent family. This is a case in
which straightforward classical reasoning in a quantum context leads to
incorrect results. The difference between the projector $B+C$ and the pair
$\{B,C\}$ is discussed Sec.~\ref{sbct4.2}; while Sec.~\ref{sct7} shows in
detail how the reasoning process from $S_0$ and $F_4$ to ``not $C_2$'' breaks
down, and why this has important implications for the claim of counterfactual
communication.

\xb
\outl{Special $\al,\,\bt$ parameters $\ra$ 3-box paradox; $\FC_A$, $\FC_C$
  incommensurate, so no contradiction}
\xa

For the special choice of beam splitter parameters in \eqref{eqn22}, ignoring
the single framework rule leads to a paradox, Sec.~\ref{sbct3.4}: given the
same conditions, $S$ at $t_0$ and $F$ at $t_4$, the consistent family $\FC_A$
leads to the conclusion that the particle was in $A$ at $t_2$, whereas the
equally consistent family $\FC_C$ locates the particle at $t_2$ in $C$. This is
an instance of the three box paradox in quantum foundations; details of how the
CH approach resolves it (perhaps better, ``tames it'') will be found in Sec.
22.5 of \cite{Grff02c}. Here it suffices to note that $\FC_A$ and $\FC_C$ are
incompatible (to be more precise, incommensurate) families that cannot be
combined, so the contradiction that arises when using classical reasoning is
eliminated when proper quantum principles are applied.

\xb
\outl{Weak measurement analysis, Sec.~\ref{sct5}, confirms Sec.~\ref{sct3};
particle in $A$ at $t_2$}
\xa

\xb
\outl{Weak interaction sometimes $\ra$ big effect. Qubit probes in this case
are as good as Gaussian probes}
\xa

Both Vaidman \cite{Vdmn13} and Li et al.\ \cite{LAAZ13} have appealed to the
weak values produced by weak measurements to determine the particle's path. The
analysis in Sec.~\ref{sct5}, which uses qubit rather than Gaussian probes, and
employs a straightforward interpretation of the results rather than weak values
(whose connection with actual particle properties is quite obscure), supports
the conclusions reached in Sec.~\ref{sct3} using the family $\FC_A$: at the
time $t_2$ the particle was in $A$. It is worth noting that even a very weak
interaction between the probe and the measured system can on rare occasions
produce very large perturbations of the latter. And also that for the fairly
simple situation considered here, qubit probes provide just as much information
as Gaussian probes, and in a form which is easier to interpret.

\xb
\outl{Ctfl commun. Vaidman critique incorrect. Salih et al.\ claim
  uses Cl reasoning, is dubious}
\xa

The comparison of the two state vector formalism and the consistent history
approach in Sec.~\ref{sct6} throws additional light on the disagreement,
mentioned in the introduction, between Vaidman and Salih et al. on the topic of
counterfactual communication. With reference to the situation in
Fig.~\ref{fgr1}, Vaidman takes the nonzero weak values for the projectors $B$
and $C$ at time $t_2$ as evidence that the particle is present in both those
channels. But since the weak values are neither zero nor one, the consistent
histories analysis instead regards them as evidence that the presence or
absence of the particle at these locations, conditioned on $S_0$ and $F_4$,
cannot be discussed in a consistent manner. This undermines Vaidman's criticism
of the counterfactual communication protocol on the basis that there was a
particle present where Salih et al.\ would have said there was none. But at the
same time it undermines the claim of Salih et al., as discussed in
Sec.~\ref{sct7}, that there was no particle in the $C$ channel, for that
claim relies upon classical reasoning in a quantum context in which it fails.
(To be sure, the full protocol for counterfactual communication is much more
complicated than the simple example considered in \cite{Slao14}, and we have
not attempted to analyze it. However, it seems doubtful that the full protocol
is more reliable than the simple example, at least until it is supported by a
consistent quantum mechanical analysis, and that has not yet been carried out.)

\xb
\outl{SQM lacks tools for analyzing Qm system prior to measurement. Textbooks
invoke `measurement' to avoid difficulties. CH approach should be taken
seriously.}
\xa

The contrary conclusions reached by Vaidman and by Salih et al.\ reflect the
fact that the tools needed to analyze events in a microscopic quantum system
prior to a macroscopic measurement are not part of standard quantum mechanics,
understood as what one finds in standard textbooks. Trying to extend this by
using classical reasoning, or a phenomenology based on the two state vector
formalism, or weak measurements, can be very misleading. The success of the
calculational tools found in textbooks arises both from the fact that they
employ the quantum Hilbert space and subspaces to represent physical properties
as well as carry out calculations, and also from a judicious invocation of
``measurement,'' never properly explained, to evade the various contradictions
and paradoxes which are well known in quantum foundations. The consistent
histories approach extends standard textbook quantum mechanics in a way that
allows an analysis of microscopic quantum behavior without leading to
contradictions and insoluble paradoxes. At the same time it provides a
fully quantum mechanical description of the measurement process, including what
it is that is measured, thus getting rid of the measurement problem. While
consistent histories may not be the final word in quantum interpretation, it
deserves to be taken very seriously in the absence of alternatives which can
provide a plausible description of the microscopic quantum properties and
processes that precede measurements \cite{Grff15}.

\end{document}